\documentclass[fleqn,twoside]{article}

\usepackage{epsfig}
\usepackage{cite}

\usepackage{graphicx}
\usepackage[figuresright]{rotating}

\usepackage{espcrc2}
\usepackage{amsmath}

\title{Foam: A General purpose Monte Carlo Cellular Algorithm%
\thanks{Work partly supported 
  by the European Community's Human Potential
  Programme under contract HPRN-CT-2000-00149 ``Physics at Colliders'',
  by Polish Government grants
  KBN 5P03B09320 
  and 2P03B00122, 
  and by NATO grant PST.CLG.977751.
  }}

\author{
  S. Jadach%
  \address{Henryk Niewodniczanski Institute of Nuclear Physics,\\
    ul. Radzikowskiego 152,  31-342 Cracow, Poland 
    }}
      
\begin{document}

\begin{abstract}
A general-purpose, self-adapting Monte Carlo (MC) algorithm
implemented in the program {\tt Foam} is described.
The high efficiency of the MC, that is small maximum weight or variance
of the MC weight is achieved by means
of dividing the integration domain into small cells.
The cells can be $n$-dimensional simplices,
hyperrectangles or a Cartesian product of them.
The grid of cells, ``foam'', is produced in the process of the binary
split of the cells.
The next cell to be divided and the position/direction
of the division hyperplane
is chosen by the algorithm which optimizes
the ratio of the maximum weight to the average weight or (optionally) the total variance.
The algorithm is able to deal, in principle, with an arbitrary pattern
of the singularities in the distribution.
\end{abstract}
\maketitle

\noindent
{\bf Introduction and motivation}\\
For the problem of function minimalization of the arbitrary user-function
one applies MINUIT or other program from the NAG library.
One finds also many {\em general purpose} programs  for integration
of ``arbitrary'' integrand.
The above {\em general-purpose} tools work, in principle, 
for a very wide range of the user-functions.
For {\em multi-dimensional Monte Carlo simulation} problem,
that is for the problem of generating randomly points according to
$n$-dimensional distribution, there is precious little 
examples of the {\em general purpose} Monte Carlo Simulators (GPMCS),
that is programs which work (in principle) for an arbitrary integrand.

Inevitably the GPMCS has to work in 2 stages: ``exploration'' and ``generation''.
During exploration GPMCS is ``digesting''
the entire shape of the $n$-dimensional distribution $\rho(x_1,x_2,...x_n)$ to be generated
and memorizes it as efficiently as possible using all CPU power and memory available.
Obviously, for the memorized $\rho'(x_1,x_2,...x_n)$
a method of the MC generation of the points $\vec x$
{\em exactly} according to $\rho'(\vec x)$, has to be available.
The quality of the distribution of the weight $w=\rho/\rho'$
for events in the {\em generation}
(small variance, good ration of maximum to average, etc.)
is determined by the algorithm of the {\em exploration}.
In other words, the quality of the ``target weight distribution''
in the latter generation is
determining/driving the algorithm of the former exploration.

At the time of writing up this contribution a detailed description of the algorithm
and the program is available in ref.~\cite{foam2:2002}.
The original, less advanced version was described in ref.~\cite{foam1:2000}.
The other old and new efforts in the area of general purpose MC
methods see 
refs.~\cite{Lepage:1978sw,Kawabata:1995th,Ohl:1998jn,Manankova:1995xe,%
Doncker:1998,Doncker:1999,Doncker:parint1}

\begin{figure*}[!ht]
\centering
\setlength{\unitlength}{0.05mm}
\begin{picture}(1600,1100)
\put( 450,1050){\makebox(0,0)[b]{\large (a)}}
\put(1100,1050){\makebox(0,0)[b]{\large (b)}}
\put(1850,1050){\makebox(0,0)[b]{\large (c)}}
\put( -200, 550){\makebox(0,0)[lb]{\epsfig{file=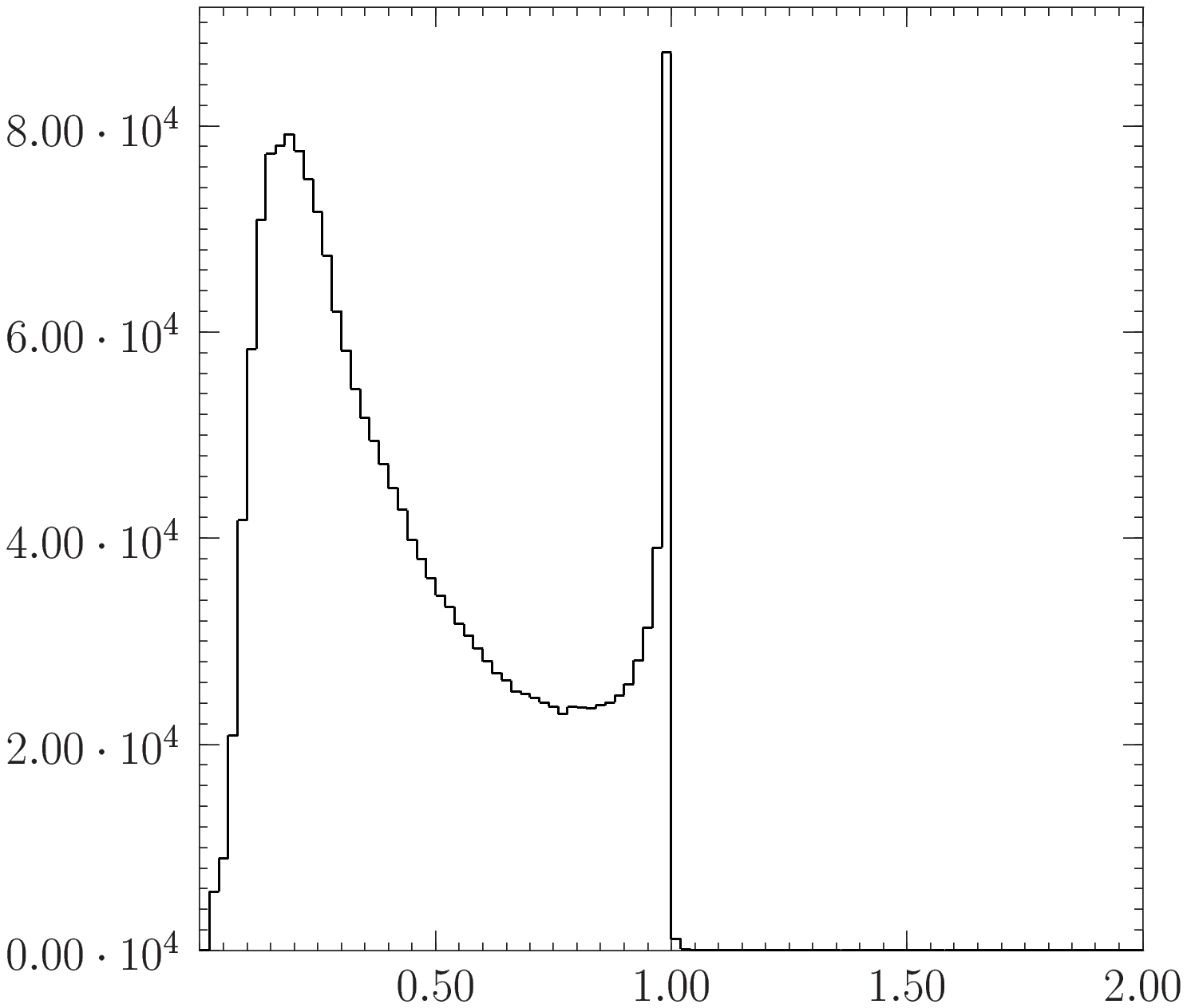,width=38mm}}}
\put(  550, 550){\makebox(0,0)[lb]{\epsfig{file=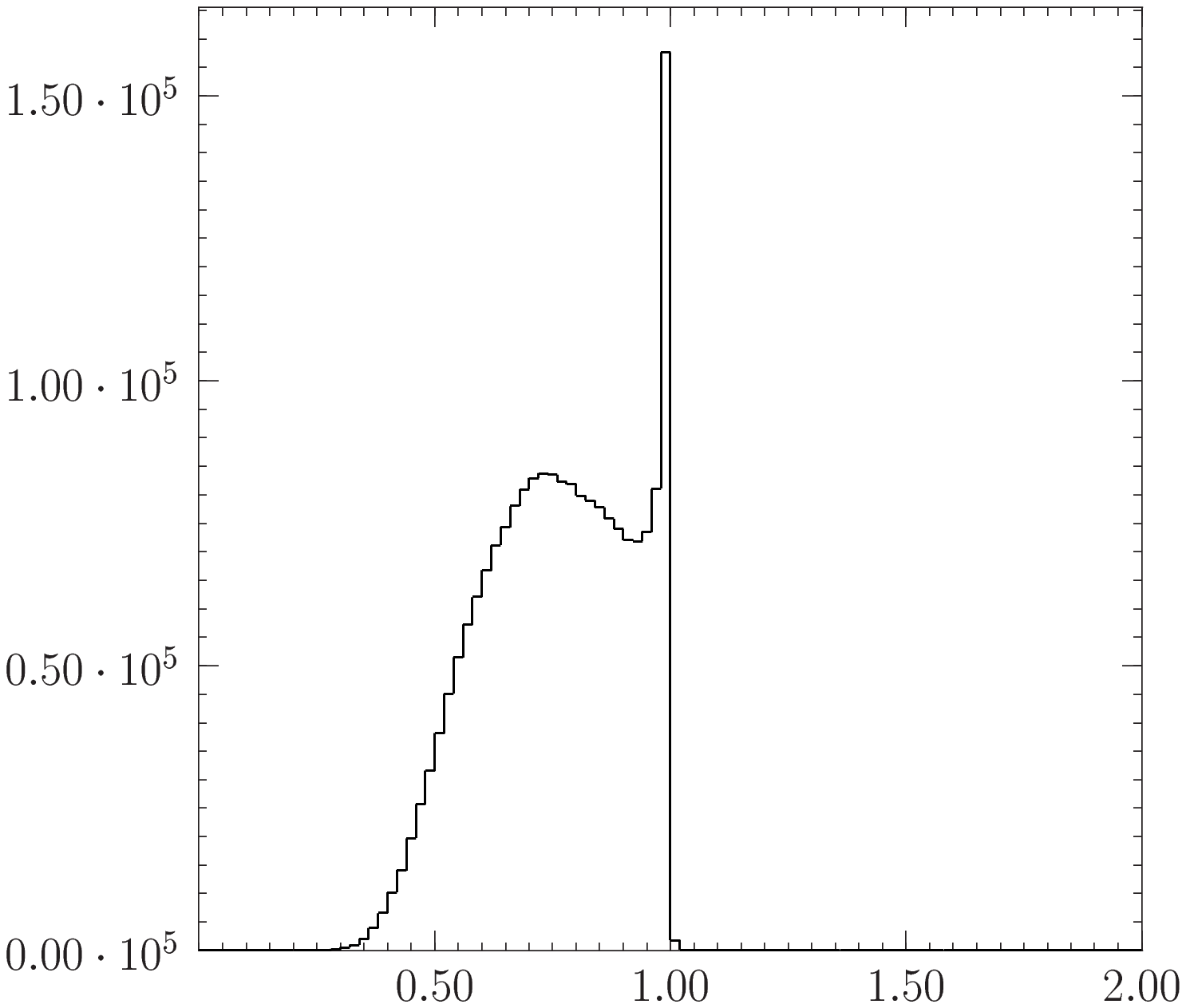,width=38mm}}}
\put( 1300, 550){\makebox(0,0)[lb]{\epsfig{file=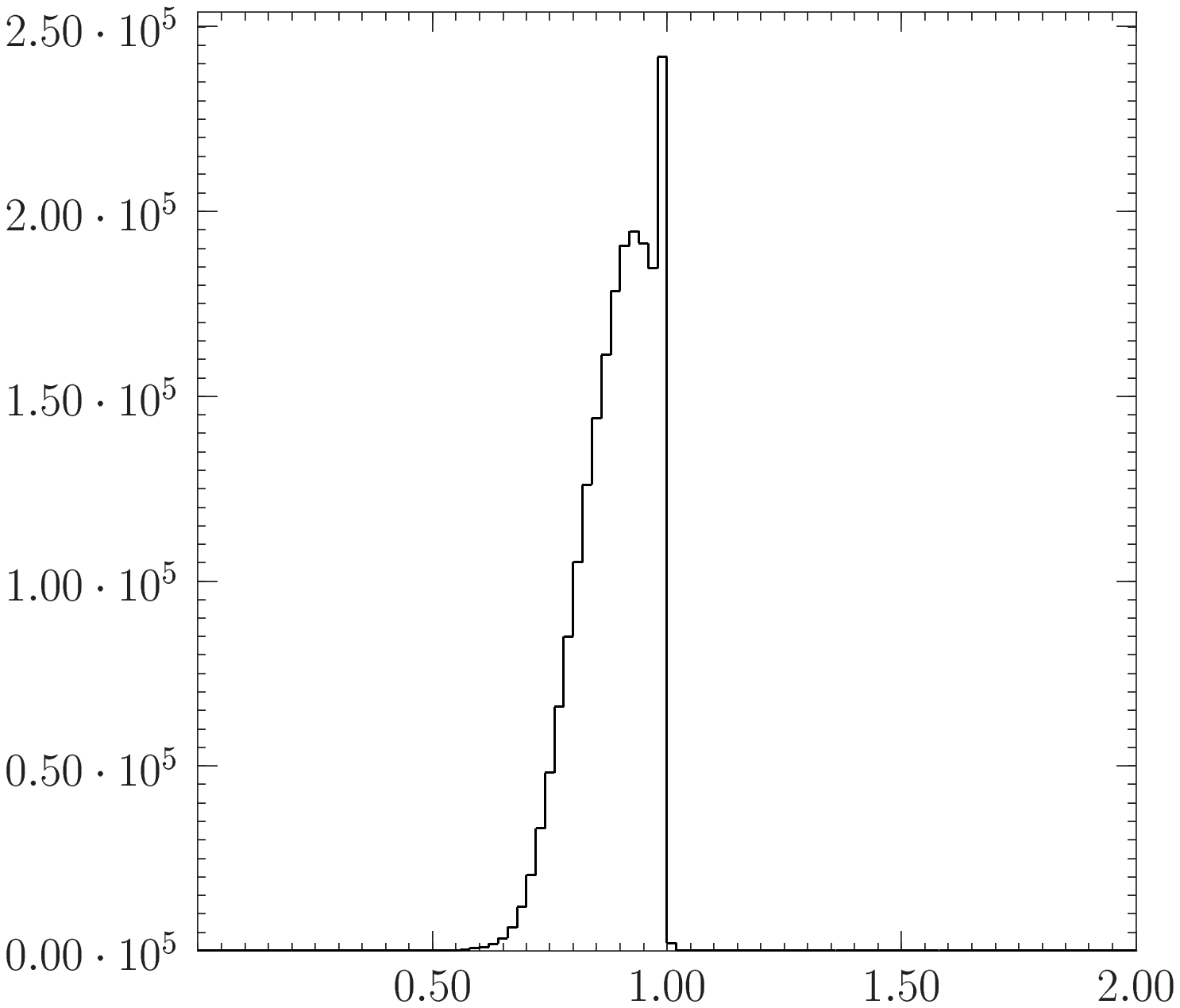,width=38mm}}}
\put( 450,400){\makebox(0,0)[b]{\large (d)}}
\put(1100,400){\makebox(0,0)[b]{\large (e)}}
\put(1850,400){\makebox(0,0)[b]{\large (f)}}
\put( -200, -100){\makebox(0,0)[lb]{\epsfig{file=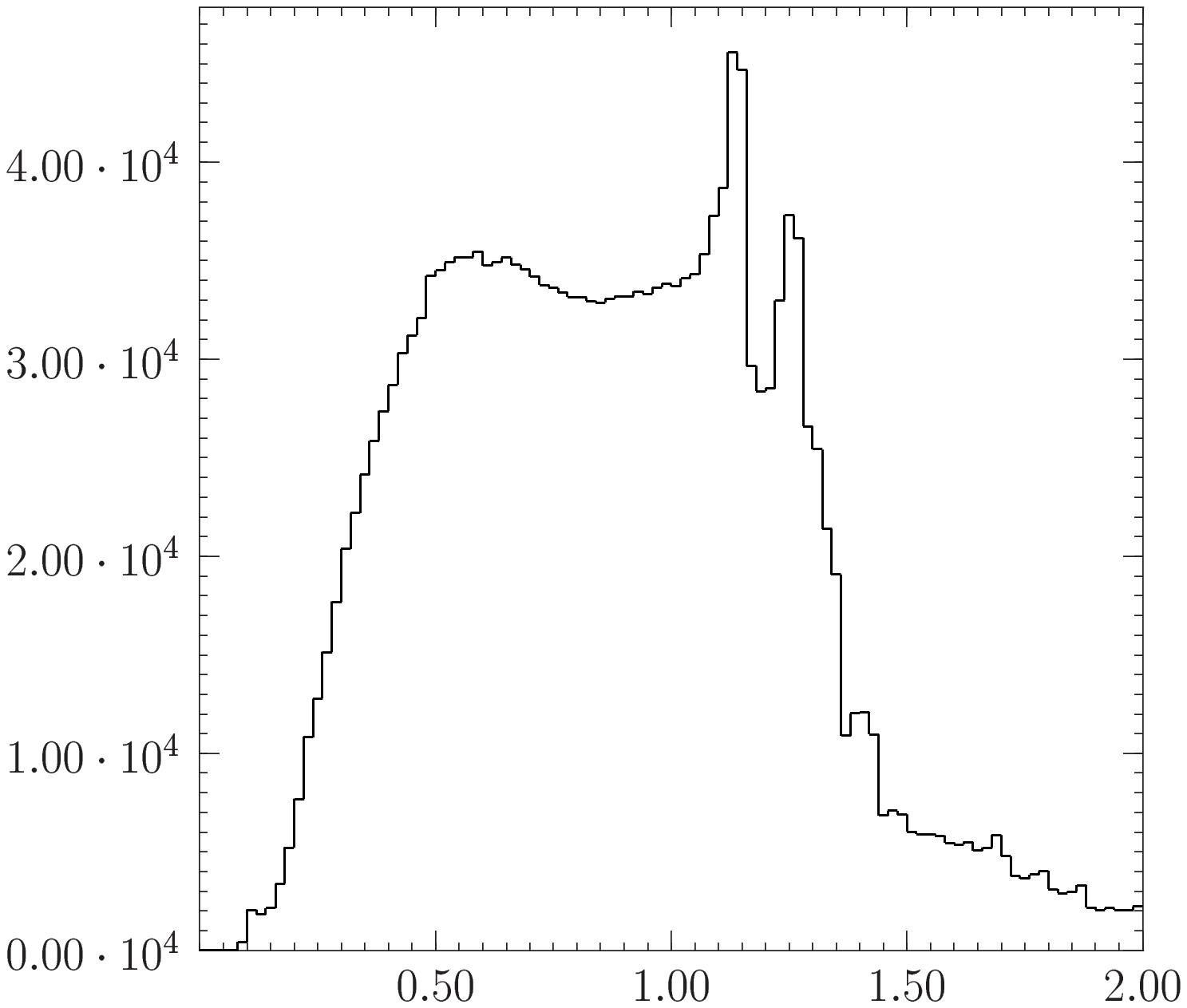,width=38mm}}}
\put(  550, -100){\makebox(0,0)[lb]{\epsfig{file=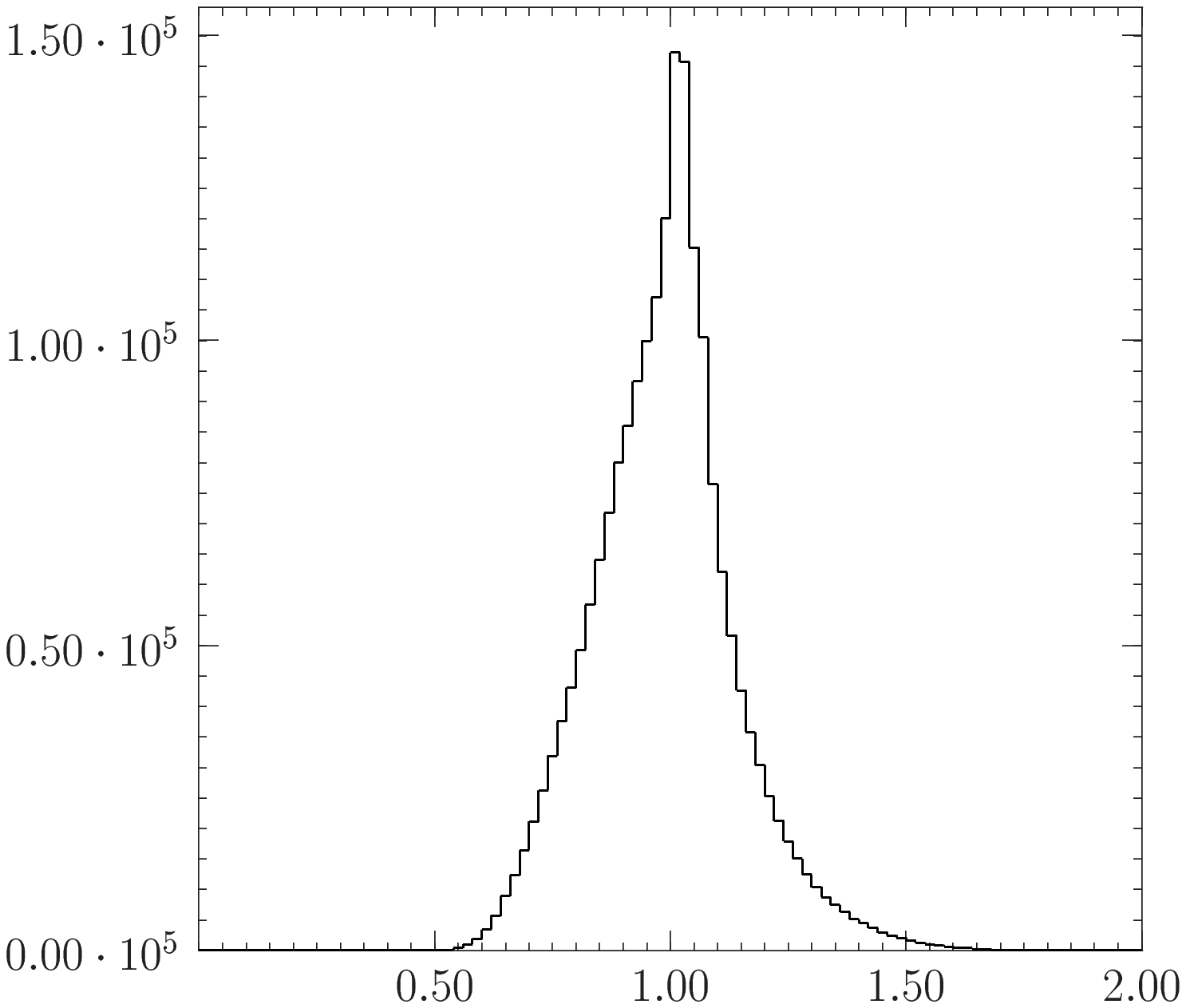,width=38mm}}}
\put( 1300, -100){\makebox(0,0)[lb]{\epsfig{file=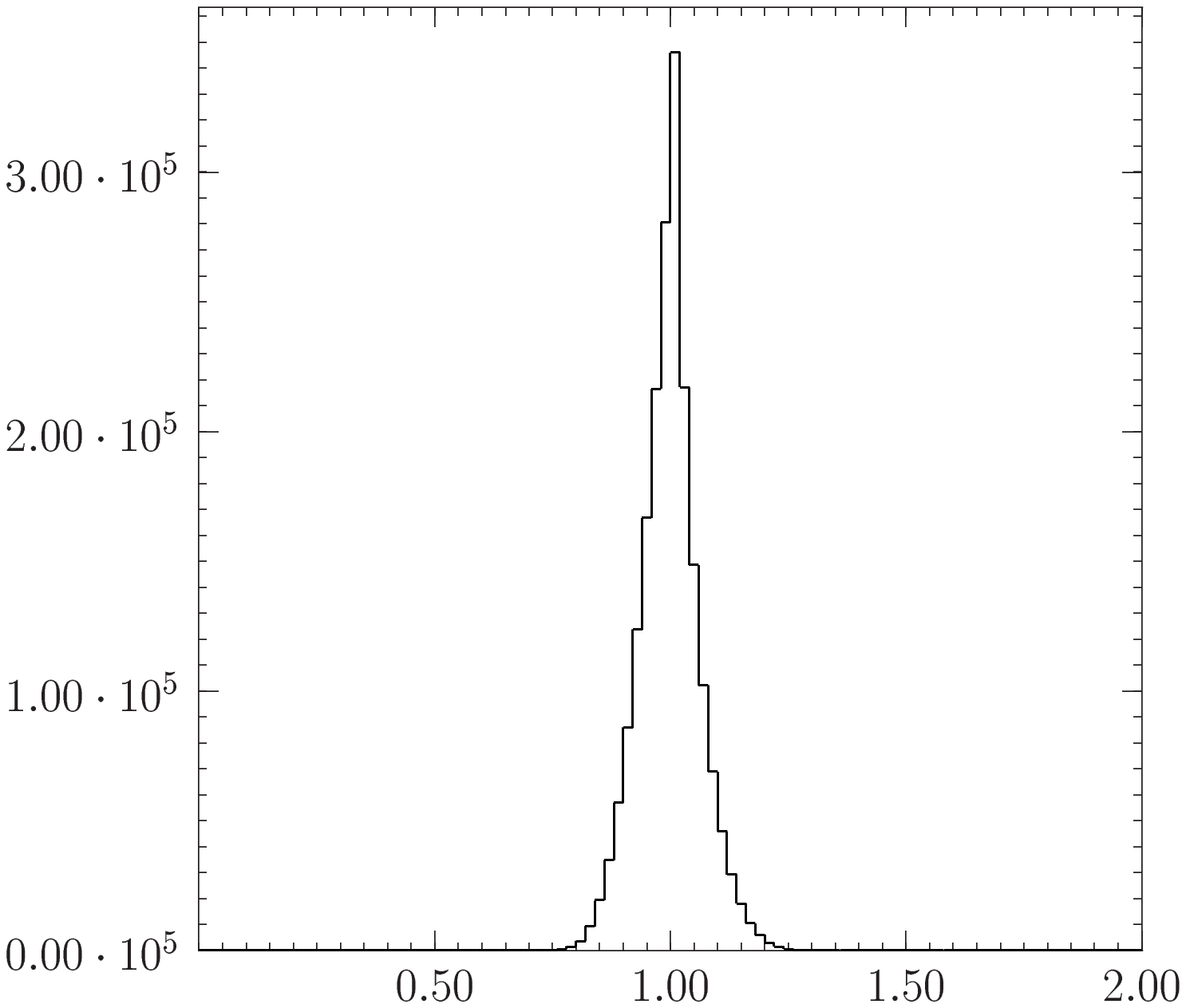,width=38mm}}}
\end{picture}
\caption{\small\sf
  Weight distribution of the {\tt Foam} with the maximum weight optimization
  (a-c) compared to the variance optimization (d-f).
  Number of cells is 200, 2000 and 20000 correspondingly.
}
\label{fig:wtdist1}
\end{figure*}

\noindent
{\bf Cellular algorithm of the {\tt Foam}}\\
The most obvious method to minimize
the variance (or maximum weight) of the target weight distribution in generation
is to split integration domain into many cells%
\footnote{This general idea is at least 40 years old.},
such that $\rho(\vec x)$ is approximated by constant $\rho'(\vec x)$ within each cell.
This is a {\em cellular class} of general purpose MC algorithms%
\footnote{``Stratified sampling'', used in the literature, has a narrower meaning.}
which always involves a kind of cellular exploration of the user defined distribution.

The obvious questions are:
what kind (shape) of cells to use and how to cover the space with cells?
In {\tt Foam} algorithm/program three types of cells are used:
pure {\em simplices}, pure {\em hyperrectangles} and {\em Cartesian products of them}.
Cells of this kind can be parametrized rather easily and the related
computer memory consumption is not excessive.

The system of cells can be created all at once (like in VEGAS~\cite{Lepage:1978sw})
or in an evolutionary way, by the ``splitting process''.
The {\tt Foam} algorithm does the {\em binary split} of cells.
Next cell to be split is chosen by examining the ``target weight distribution'',
that is of the event generation in the latter phase.
The binary split provides automatically the {\em full coverage} of the space,
simply because the primary ``root cell'' is identical with the entire integration domain
and there is never any mismatch between parent cell and two daughter cells.

Another important issue is the question of the variance reduction 
versus maximum weight reduction.
In construction of the {\tt Foam} algorithm we have put most effort on
the minimization of the ratio of the maximum weight to the average weight
$w_{\max}/\langle w \rangle$.
This parameter is essential, if we want to transform
$w$-ted events into $w=1$ events, at the latter stage of the MC generation.
Minimizing maximum weight is not the same as minimizing variance
$\sigma=\sqrt{ \langle w^2 \rangle -\langle w \rangle^2 }$.
Usually minimizing $w_{\max}$ is more difficult.
In {\tt Foam}  minimizing variance is also implemented and optionally available.
It can be useful in the case when a MC with $w$-ted events is acceptable.
In fig.~\ref{fig:wtdist1} we show two examples of the weight distribution evolution
in the {\tt Foam}, when adding more and more cells.

In the Foam algorithm
we define two auxiliary distributions $\rho'(x)$ and $\rho_{loss}(x)$
related to integrand $\rho(x)$.
Both are constructed together step by step,
simultaneously with the construction of the foam of cells,
in the exploration process.
When our ultimate aim is to minimize $w_{\max}$, we define
\begin{displaymath}
  \begin{split}
    \rho'(x)&\equiv \max_{y\in Cell_I} \rho(y),\quad \hbox{for}\quad x\in Cell_I,\\
    R_{loss}&= \int d^nx\; [\rho'(x) -\rho(x) ]=\int d^nx\;\rho_{loss}(x).
  \end{split}
\end{displaymath}
here $\rho_{loss}$ is the difference between 
the ``ceiling distribution'' $\rho'$ and the actual distribution $\rho$
from which it is derived.
The rejection rate in the final MC run is just proportional
to $\int\rho_{loss}(x)$ by construction, i.e. the rejection rate $= R_{loss}/R$.
Also, $\rho_{loss}(x)$ has a clear geometrical meaning, see below.
When our principal aim is to
minimize the ratio of the variance to the average of the weight, $\sigma/\langle w \rangle$,
in the final MC generation, we are led to the following definition:
\begin{displaymath}
  \begin{split}
    &\rho'(x) \equiv \sqrt{ \langle \rho^2 \rangle_I},\quad \hbox{for}\quad x\in Cell_I,\\
    &\rho_{loss}(x) \equiv \sqrt{ \langle \rho^2 \rangle_I} -\langle \rho \rangle_I,
    \quad \hbox{for}\quad x\in Cell_I.
  \end{split}
\end{displaymath}
The average $\langle ... \rangle_I$ is over the $I$-th cell%
\footnote{See ref.~\cite{foam2:2002} for a detailed derivation of the above prescription.}.
The ratio $\sigma/\langle w \rangle$ in the final MC generation
is a monotonous ascending function of $R_{loss}=\int \rho_{loss}(x) dx^n$ --
hence, minimization of $R_{loss}$ is equivalent to minimization of $\sigma/\langle w \rangle$.

\begin{figure*}[!ht]
\begin{center}
{\epsfig{file=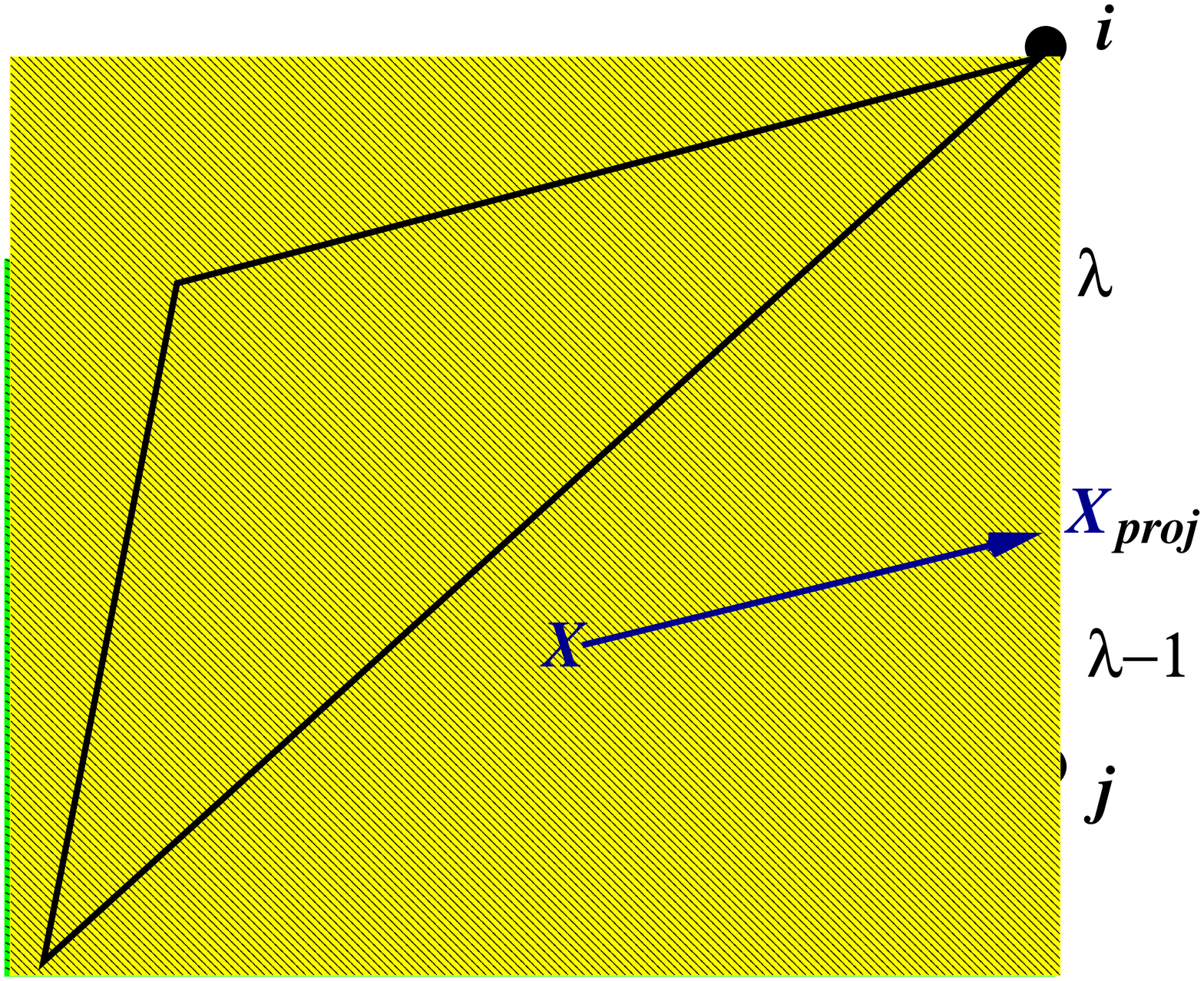,height=55mm,angle=270}}
\vspace{-5mm}
{\epsfig{file=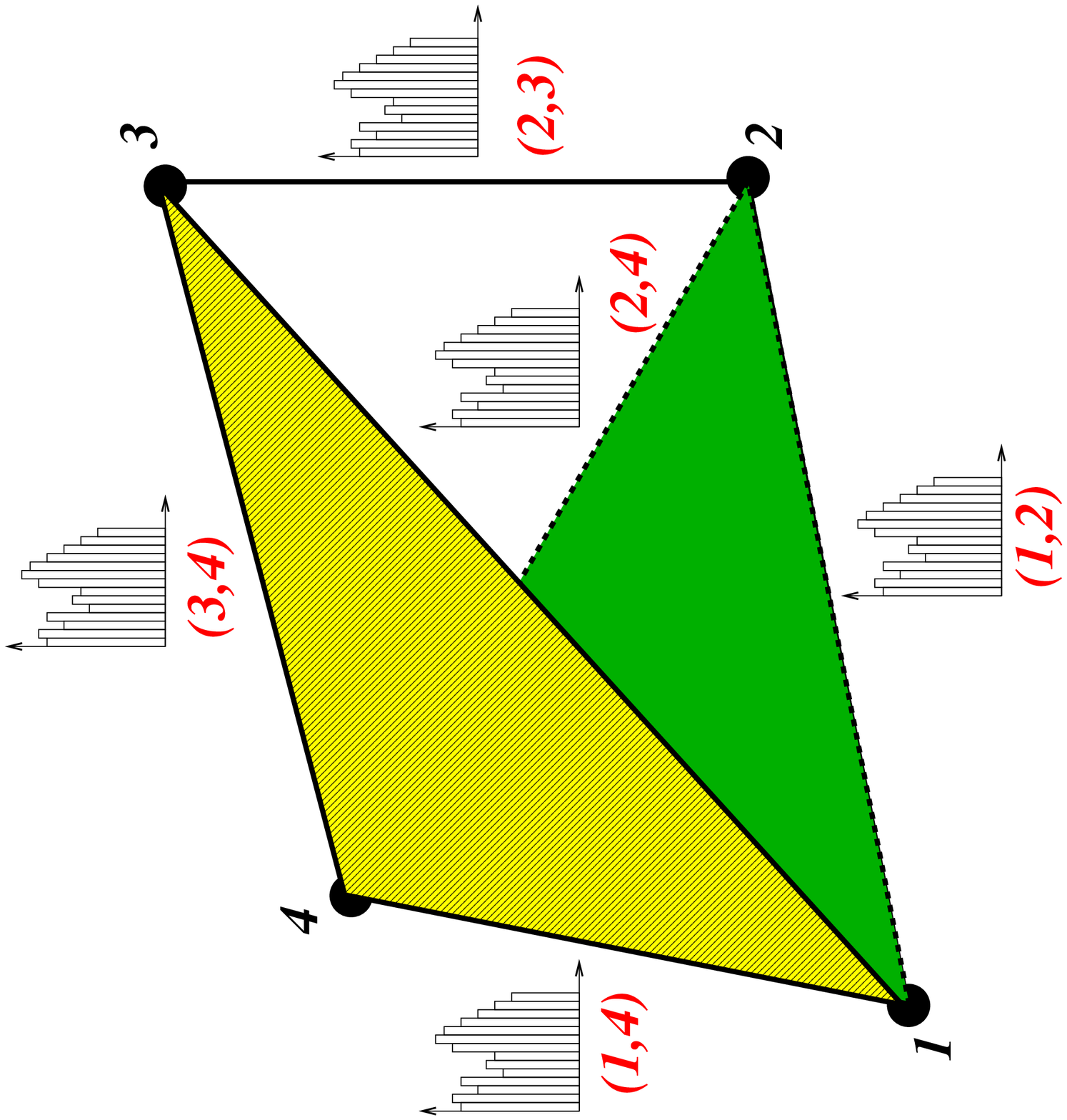,height=60mm,angle=270}}
\end{center}
\caption{\small\sf
  Geometry of the split of the 3-dimensional simplex cell.
}
\label{fig:split_simplex}
\end{figure*}

The basic  rules governing binary split of a cell is that
each split of a cell: $\omega \to \omega'+\omega''$ should decrease the total $R_{loss}$
and the decrease should be as big as possible.
How to get the best total decrease $\Delta R_{loss}$?\\
(1) For each next cell split we choose a cell with the biggest $R_{loss}$.\\
(2) Position/direction of a plane dividing a parent cell into two daughter cells
  is chosen to get the smallest total $R_{loss}$.\\
The actual procedure relies on the small MC exercise done for each single cell
during its exploration, in which events are generated with the flat distribution,
weighted with $\rho$ and projected onto $n$ (simplical case) or
$n(n+1)/2$ (hyperrectangular case) of the cells.
Resulting projection histograms are then analysed
and the best ``division geometry'' is found,
for which the estimate of $\Delta R_{loss}$ is calculated.
How the geometry of the division algorithm looks like?
In case of a $n$-dimensional simplex defined by $n+1$ vertices,
a pair of vertices $x_i$ and $x_j$ is chosen
and a new vertex $Y$ is put somewhere on the line in between:
$ Y=\lambda x_i+(1-\lambda)x_j,\;\; 0<\lambda<1$.
Two daughter simplices are defined with two list of vertices:
$(x_1,...,x_{i-1},Y,x_{i+1},...,x_{j-1},x_j,x_{j+1},...,x_n,x_{n+1})$
and
$(x_1,...,x_{i-1},x_{i+1},...,x_{j-1},Y,x_{j+1},...,x_n,x_{n+1})$.
Now, how do we choose a pair $(i,j)$ and the value of $\lambda$?
A short sample of the MC events (100-1000) is generated $\in$ cell.
Each MC point is projected (see fig.~\ref{fig:split_simplex})
$X\to Y$ onto an edge $(i,j), i\neq j$.
For each $(i,j)$ the $dN/d\lambda$ is histogrammed,
and its ``loss'' functional $R_{loss}$ is estimated.
The optimal edge $(i,j)$ with the biggest loss is selected.
For the optimal edge the cell division ration $\lambda$ is deduced from its histogram
($\lambda$ is always a rational number, $n/N_{bin}$).

\begin{figure}[!ht]
\begin{center}
\epsfig{file=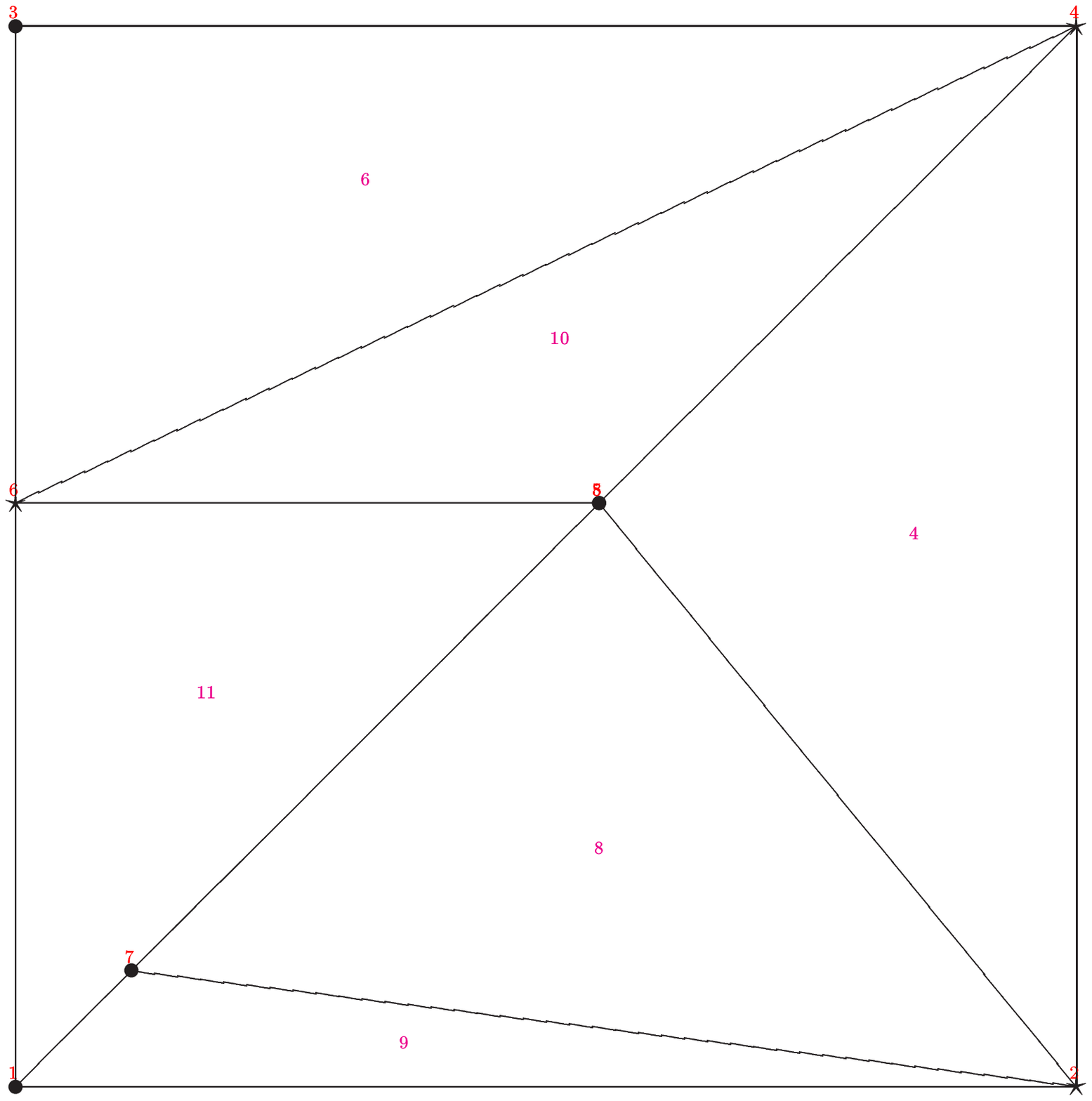,width=24mm}
\epsfig{file=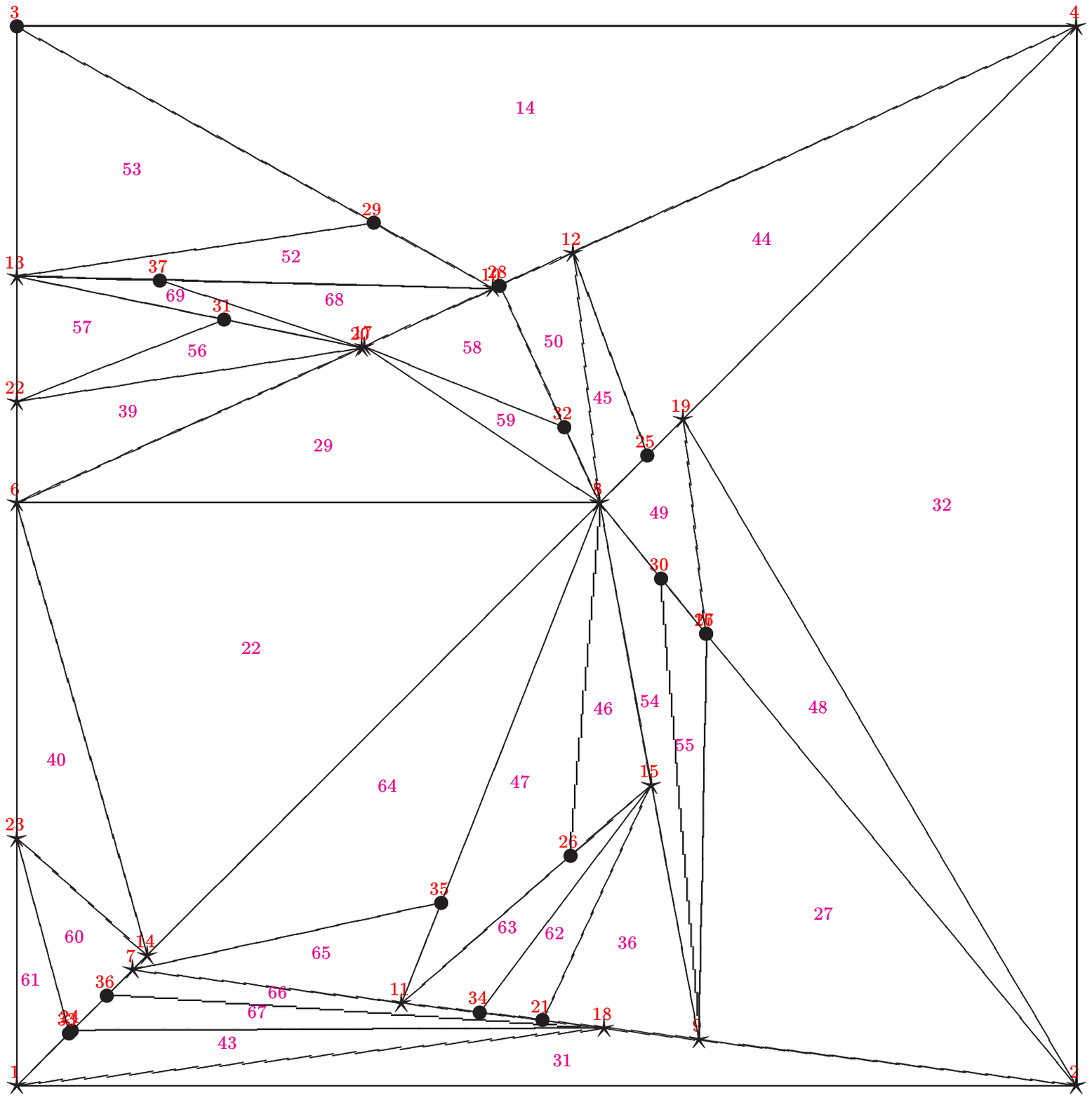,width=24mm}
\epsfig{file=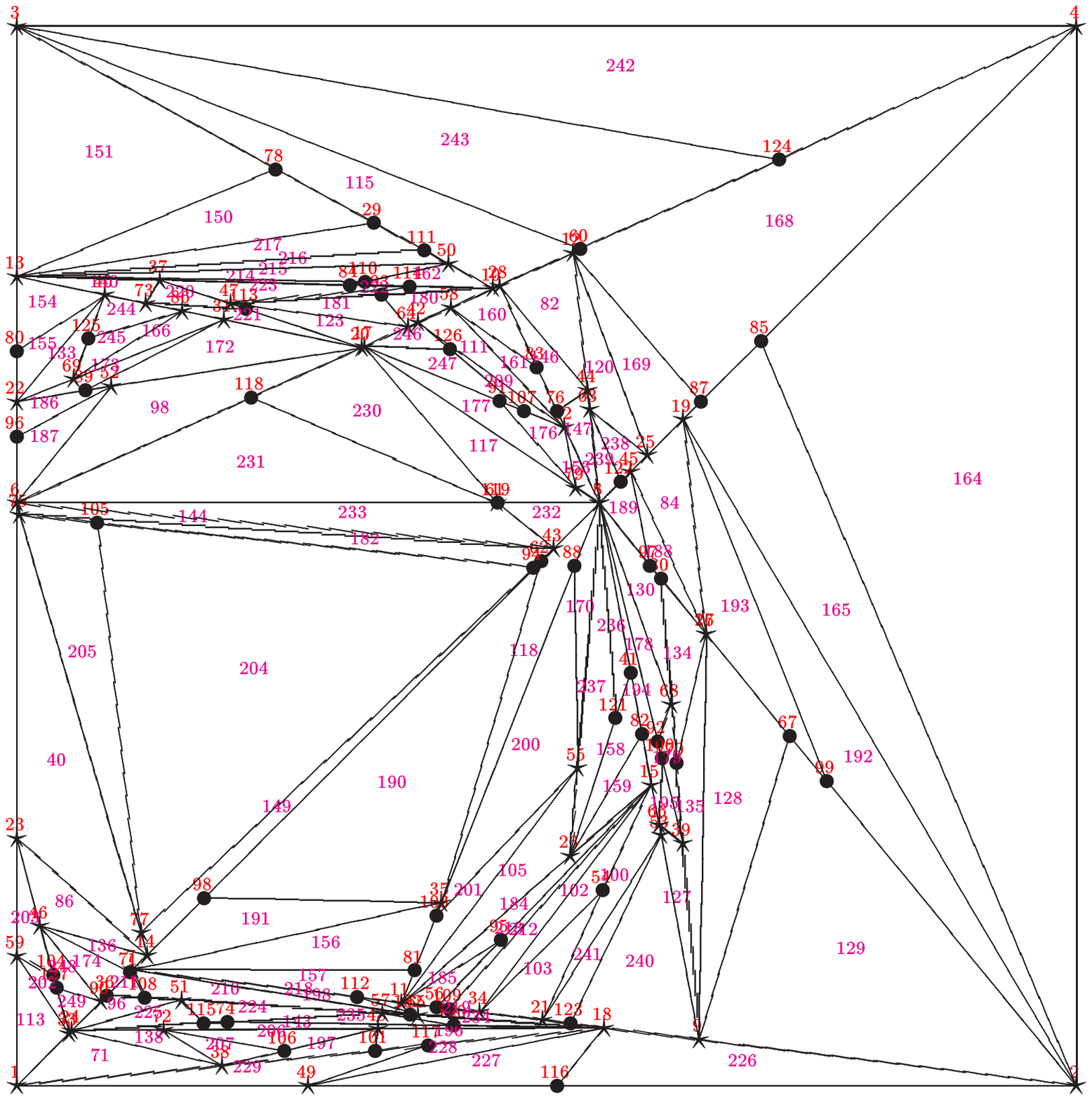,width=24mm}\\
\epsfig{file=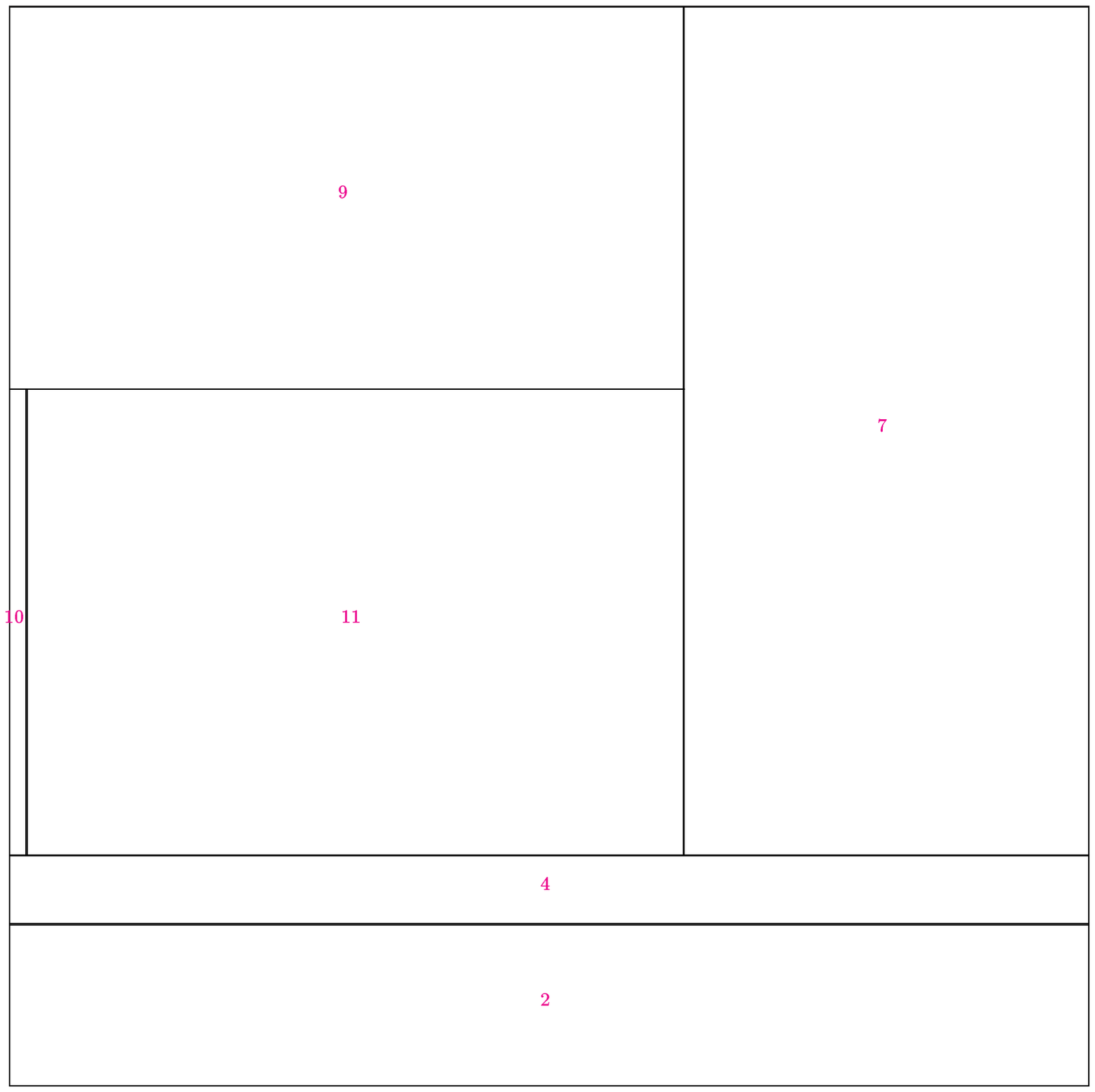,width=24mm}
\epsfig{file=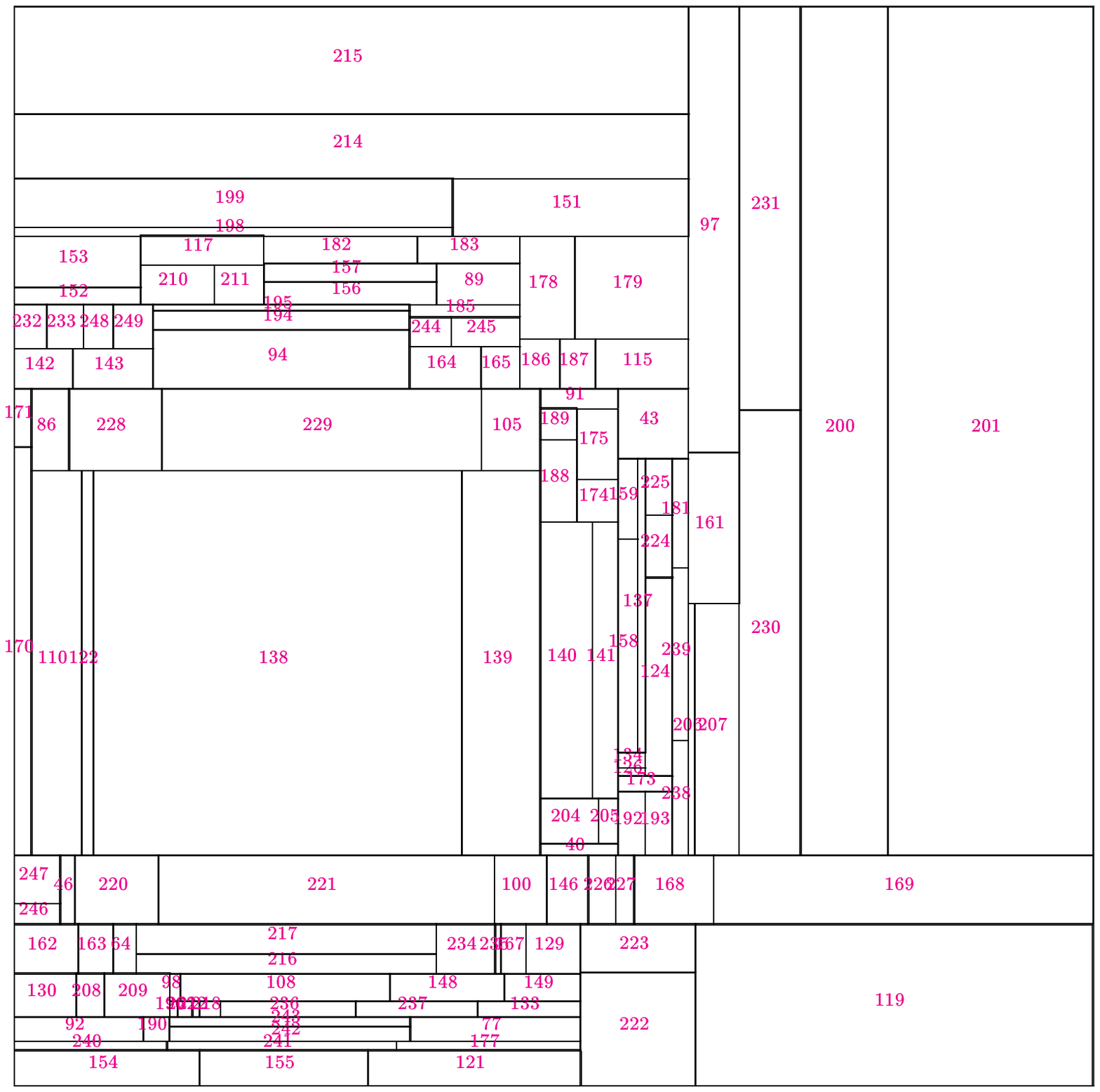,width=24mm}
\epsfig{file=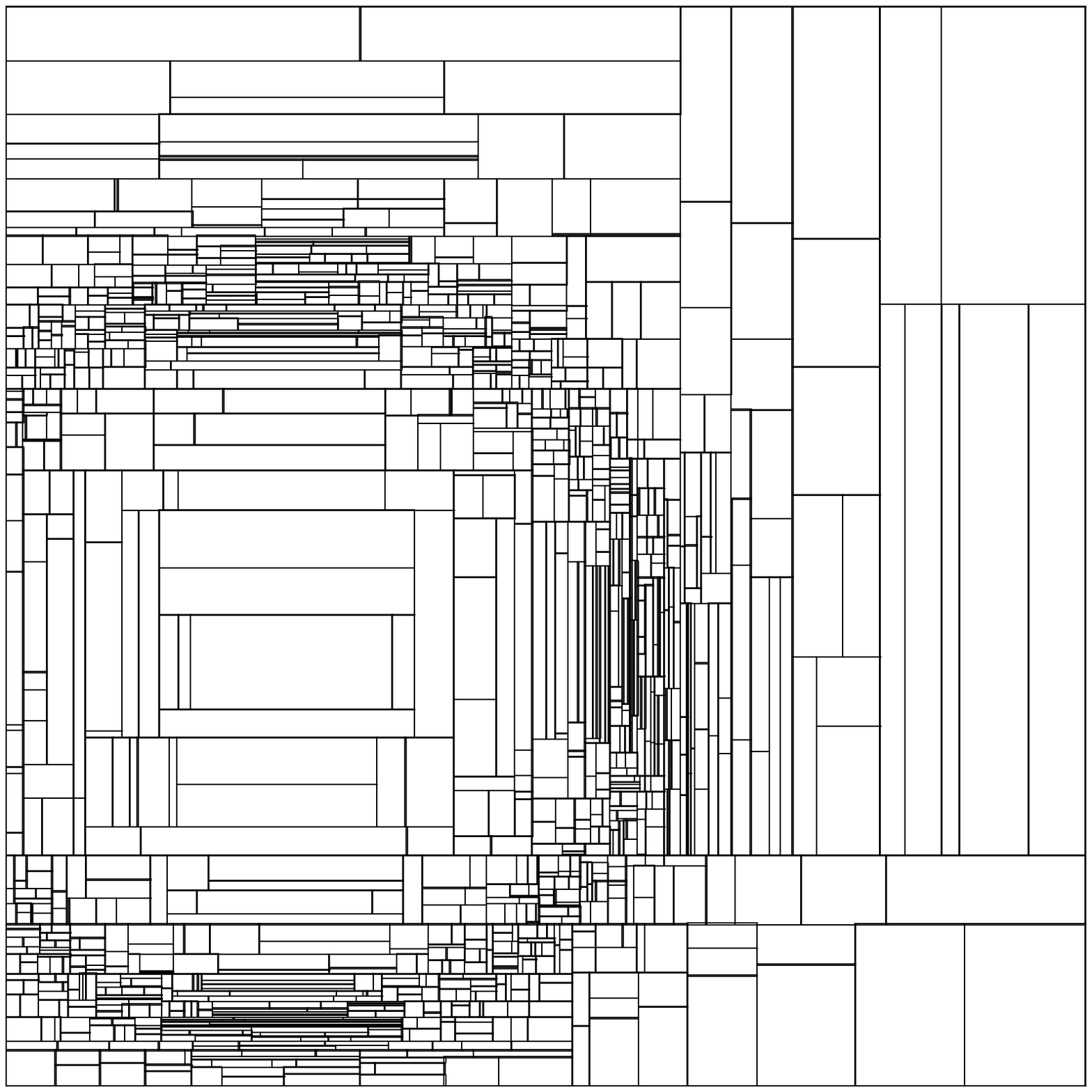,width=24mm}\\
\epsfig{file=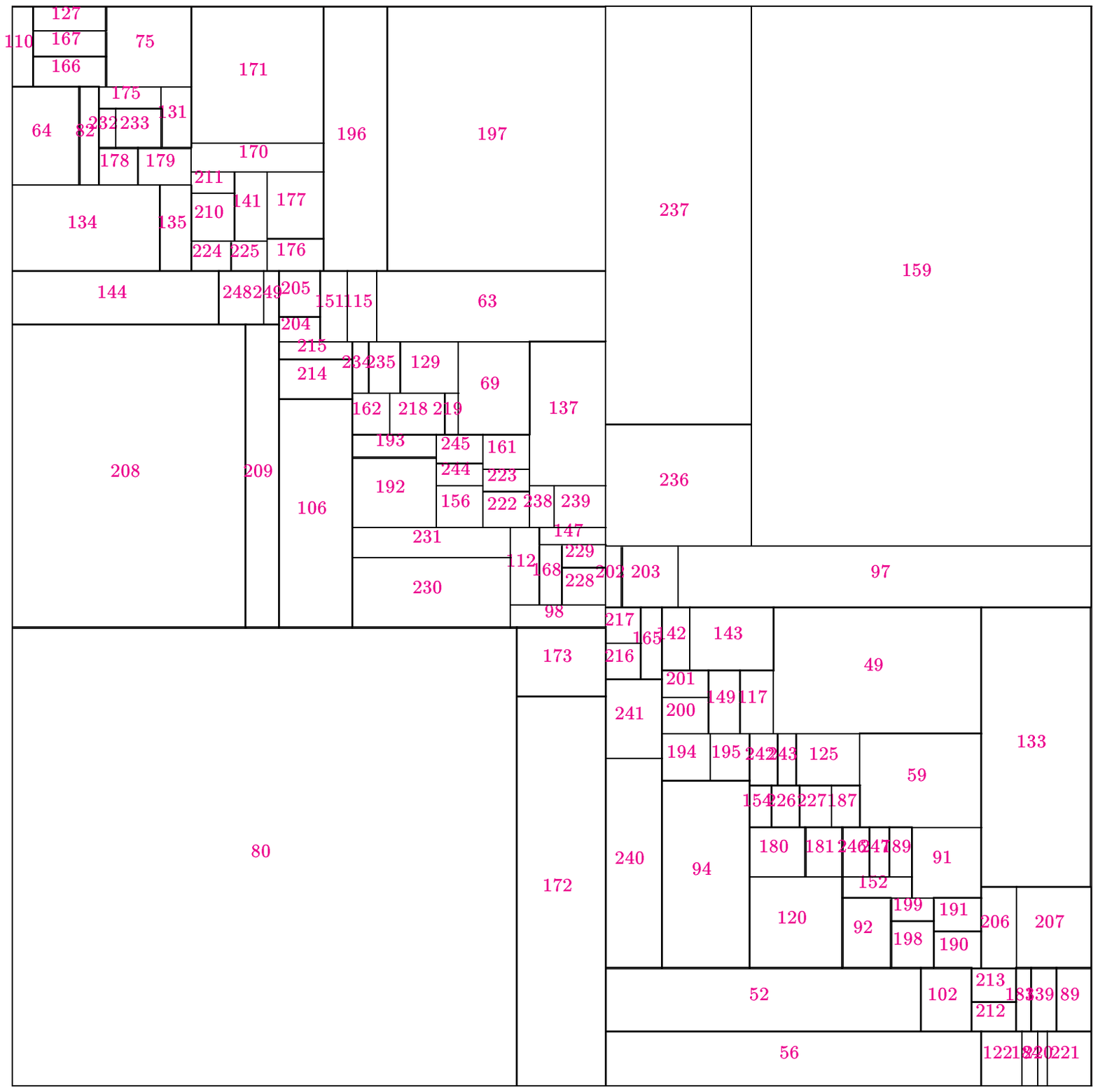,width=24mm}
\epsfig{file=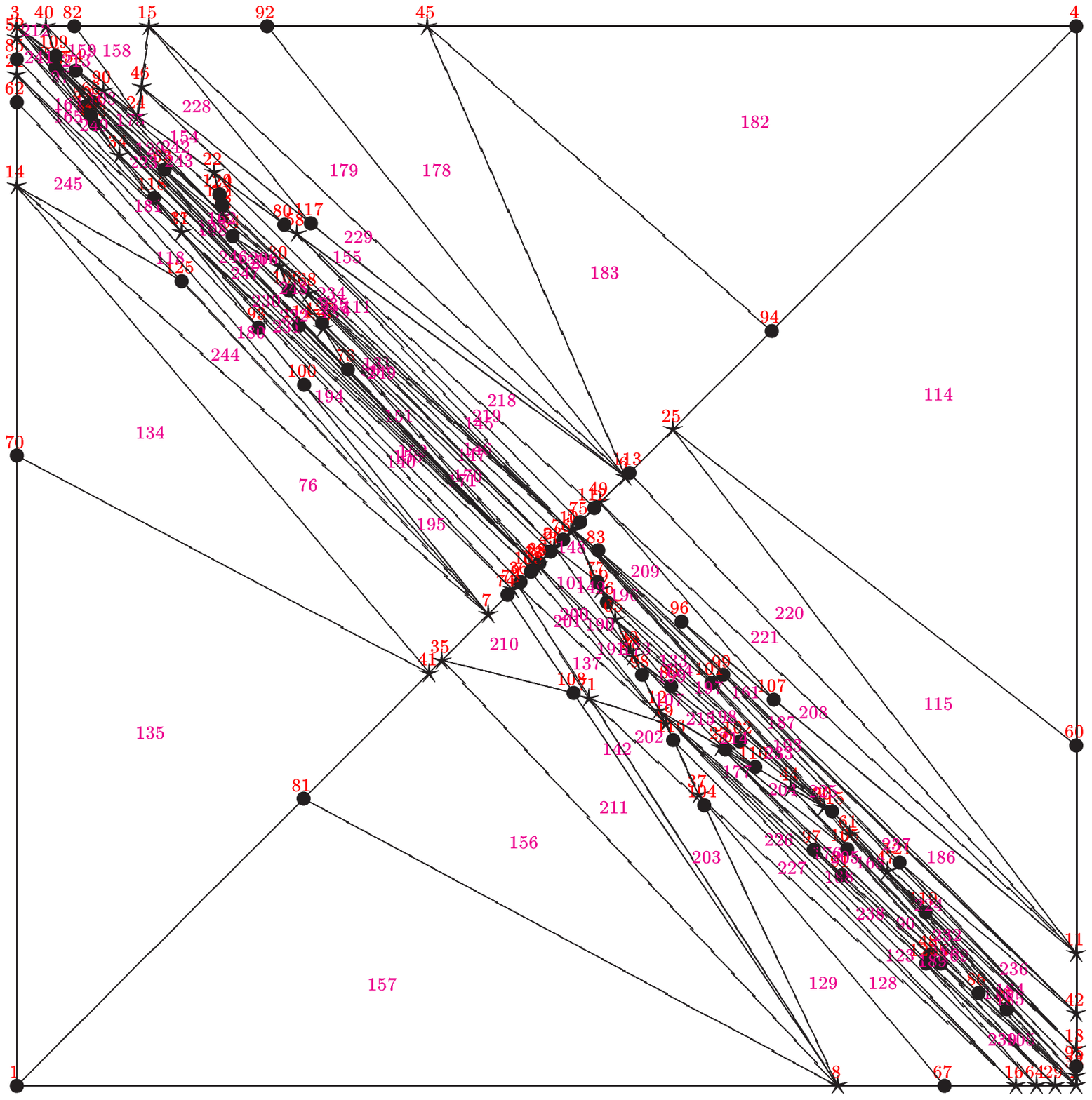,width=24mm}
\epsfig{file=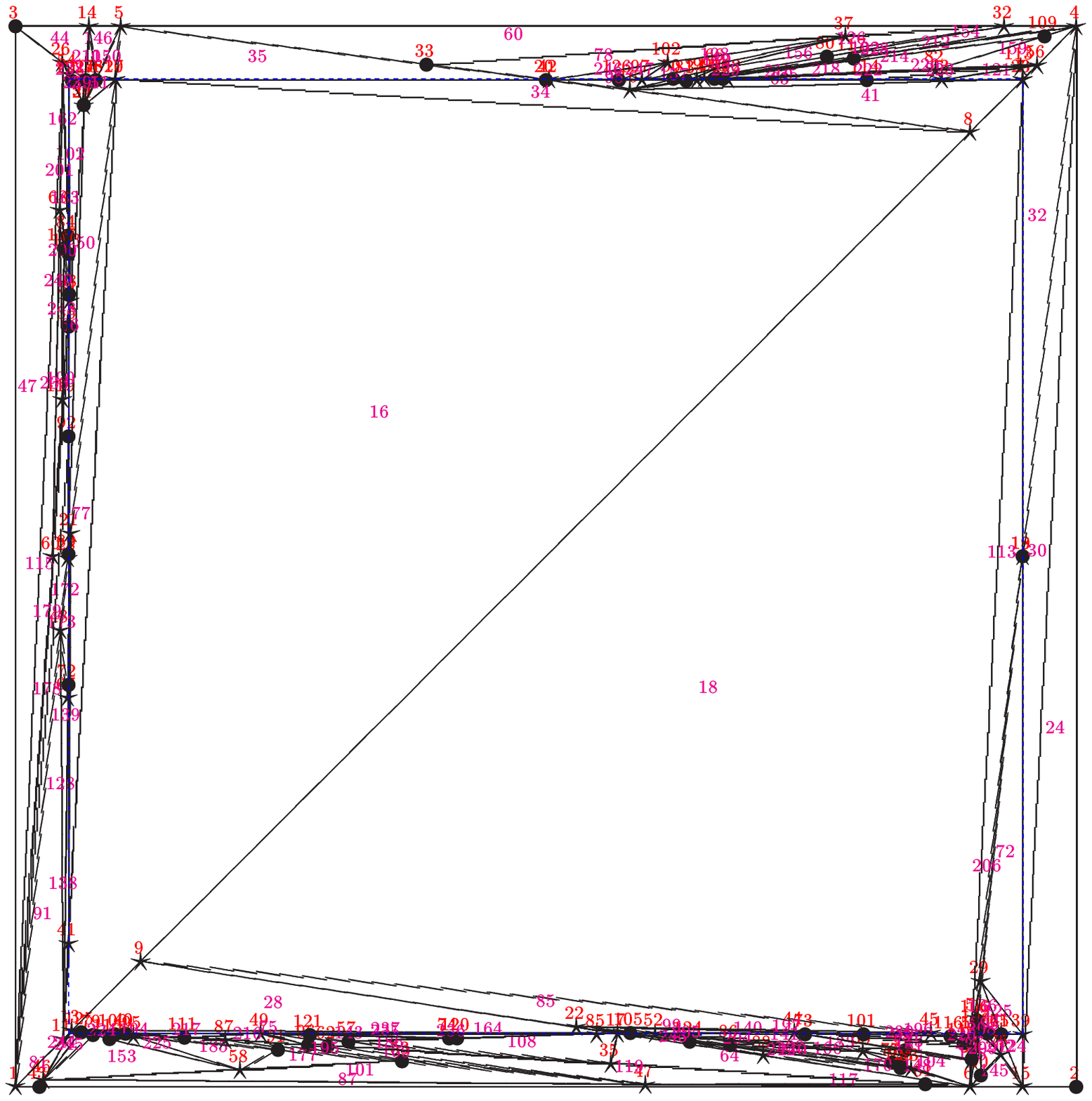,width=24mm}
\vspace{-5mm}
\end{center}
\caption{\small\sf
  Examples of the 2-dimensional foam. Number of cells from 10 to 2500.
}
\vspace{-5mm}
\label{fig:evolution}
\end{figure}

In fig.~\ref{fig:evolution} we show several examples of the evolution of the
2-dimensional foam of cells in case of simplices (triangles) and rectangles for
up to 2500 cells.

The important practical question is whether hyperrectangular cells or simplical cells are beter
as a building blocks. There is no simple answer.
Simplices are limited to low dimensions $n<6$ because calculation of increasingly large number
of determinants leads to much CPU time consumption.
In addition, for simplices memory consumption is $\sim$ 16$n$~Bytes/Cell
while for hyperrectangles memory consumption can be limited to below 50Bytes/Cell
{\em independently} of $n$. See ref.~\cite{foam2:2002} for details.
Experimenting with many testing functions has also shown that quite often
hyperrectangles provide final MC efficiency 
as good or sometimes even better than for simplices.

The method of cell division relying on the analysis of the projection histograms
requires many function calls in the cell exploration.
The overall efficiency of the final MC is best for large number of cells.
That points towards necessity of the large number of the function calls during 
the exploration, which may lead to excessive CPU time consumption.
The following efficient trick was found to limit number of the function calls
in the cell exploration:
During MC exploration of a given new cell we continuously monitor 
the number of accumulated effective $W=1$ events,
$N_{eff} = { (\sum w_i)^2 \over \sum w_i^2 }$,
and stop exploration when $N_{eff}/n_{bin}>25$, where
$n_{bin}$ is the number of bins in each histogram
used to estimate the best division direction/edge and parameter.
In this way the increase of $N_{samp}$ is not wasted for cells inside which
the integrand is already varying very little.

\vspace{1mm}
\noindent
{\bf  Programs}\\
At present the main development version of the {\tt Foam} implementation
is available in the  C++ language.
Two other versions written in Fortran 77 are also available.
They are functionally equivalent to version in C++%
\footnote{ Future development will be restricted to C++.}.

\vspace{1mm}
\noindent
{\bf Conclusions}\\
  {\tt Foam} algorithm is a versatile {\em adaptive, general-purpose} 
  type of an algorithm based on the {\em cellular division} of the integration domain.
  The geometry of the ``foam of cells'' is rather simple, cells of
  {\em simplical and/or hyperrectangular} shape are constructed in the process of a binary split.
  It works in principle for arbitrary distribution --
  {\em no assumption of factorizability} as in {\tt VEGAS} of Ref.\cite{Lepage:1978sw}.
  Encoding cells  with {\em memory-efficient methods} allows up to $\sim 10^6$ cells
  to be built in the computer memory of a typical desktop computer.

%

\begin{thebibliography}{1}

\bibitem{foam2:2002}
S.~Jadach, ``{Foam: A general purpose cellular Monte Carlo event generator},''
\href{http://arXiv.org/abs/physics/0203033}{{\tt physics/0203033}}.

\bibitem{foam1:2000}
S.~Jadach, ``Foam: Multi-dimensional general purpose Monte Carlo generator with
  self-adapting symplectic grid,'' {\em Comput. Phys. Commun.} {\bf 130} (2000)
  244--259,
\href{http://arXiv.org/abs/physics/9910004}{{\tt physics/9910004}}.

\bibitem{Lepage:1978sw}
G.~P. Lepage, ``A new algorithm for adaptive multidimensional integration,''
  {\em J. Comput. Phys.} {\bf 27} (1978)
192.

\bibitem{Kawabata:1995th}
S.~Kawabata {\em Comput. Phys. Commun.} {\bf 88} (1995) 309.

\bibitem{Ohl:1998jn}
T.~Ohl, ``Vegas revisited: Adaptive monte carlo integration beyond
  factorization,'' {\em Comput. Phys. Commun.} {\bf 120} (1999) 13--19,
\href{http://arXiv.org/abs/hep-ph/9806432}{{\tt hep-ph/9806432}}.

\bibitem{Manankova:1995xe}
G.~I. Manankova, A.~F. Tatarchenko, and F.~V. Tkachov, ``\uppercase{MILX}y way:
  How much better than \uppercase{VEGAS} can one integrate in many
  dimensions?.''
\newblock A Contribution to AINHEP-95, Pisa, Italy, Apr 3-8, 1995 (extended
  version).

\bibitem{Doncker:1998}
E.~de~Doncker and A.~Gupta, ``Multivariate integration on hypercubic and mesh
  networks,'' {\em Parallel Computing} {\bf 24} (1998) 1223--1244.

\bibitem{Doncker:1999}
E.~de~Doncker, A.~Gupta, and R.~Zanny, ``Large scale parallel numerical
  integration,'' {\em Journal of Computational and Applied Mathematics} {\bf
  112} (1999).

\bibitem{Doncker:parint1}
E.~de~Doncker, K.~Kaugars, and A.~Gupta, ``User manual of
  \uppercase{PARINT1.1}.''
\newblock http://www.cs.wmich.edu/$\sim$parint/.

\end{thebibliography}

\providecommand{\href}[2]{#2}\begingroup\raggedright\endgroup

\end{document}